\documentclass[pra,twocolumn,showpacs,amsmath,floatfix]{revtex4}
\usepackage{graphicx}
\usepackage{amssymb}

\begin{document}

\def\ket#1{|#1\rangle}
\def\bra#1{\langle#1|}
\def\av#1{\langle#1\rangle}
\def\myarrow{\mathop{\longrightarrow}}

\title{``Modes of the universe'' study of two-photon deterministic, passive quantum logical gates}

\author{Julio Gea-Banacloche}
\affiliation{Department of Physics, University of Arkansas, Fayetteville, AR 72701}
\email[]{jgeabana@uark.edu}
\author{Leno M. Pedrotti}
\affiliation{Department of Physics, University of Dayton, Dayton, OH 45469}

\date{\today}

\begin{abstract}
We use the ``modes of the universe'' approach to study a cavity-mediated two-photon logical gate recently proposed by Koshino, Ishizaka and Nakamura.  We clarify the relationship between the more commonly used input-output formalism, and that of Koshino et al., and show that some elements of this gate had been anticipated by other authors.  We conclude that their proposed gate can work both in the good and bad cavity limits, provided only that the pulses are long enough.  Our formalism allows us to estimate analytically the size of the various error terms, and to follow the spectral evolution of the field + cavity system in the course of the interaction.
\end{abstract}


\maketitle

\section{Introduction}
In a recent publication \cite{koshino}, K. Koshino, S. Ishizaka and Y. Nakamura introduced a cavity-mediated scheme to implement a deterministic photon-photon $\sqrt{\text{SWAP}}$ gate which can be operated, in principle, in a completely passive way:  no external pulses or fields are required either to initialize or read-out the atom (or equivalent three-level system) in the cavity, nor to manipulate its internal state or energy level structure in between the single-photon pulses.  This could represent a substantial simplification over existing cavity-based proposals for single-photon quantum logic, which have followed on the pioneering work \cite{DandK} of L. M. Duan and H. J. Kimble.  

Our goal in this paper is to fully characterize the conditions under which the proposal by Koshino et al. works, and to clarify its relationship to other previously-known results.  We choose do this by using the ``modes of the universe'' formalism \cite{universe0,universe}, in which the ``quasimodes'' of the optical cavity are written as superpositions of modes of a much larger cavity (the ``universe'') that encloses it.  This formalism is closely related to the one used by Koshino et al., and so we can use it to show that their results are in fact consistent with expressions obtained for related systems, over the years, by other workers making use of the simpler Collet-Gardiner input-output formalism \cite{collet}, after accounting for a non-obvious phase factor difference between the two approaches.  We derive expressions for the detuning needed to perform the $\sqrt{\text{SWAP}}$ gate that are more general than the ones in \cite{koshino}, being also valid in the good-cavity limit ($g > \kappa$, where $g$ is the atom-cavity field coupling, and $\kappa$ the bare cavity decay rate).

Our formalism allows us also to obtain simple analytical estimates for the ``error terms'' arising from the failure of various approximations, or from competing processes such as spontaneous emission, and provides a natural framework for numerical calculations.  We present results of these calculations, including the evolution of the field + cavity system in the course of the interaction.

\section{The ``modes of the universe'' formalism}

We use the formalism developed in the paper \cite{universe} (especially Appendix A), which describes a leaky cavity bounded by a perfect mirror at $z=l$ and a semitransparent mirror at $z=0$, and an auxiliary cavity (the ``outside world'') bounded by a perfect mirror at $z=-L$ ($L\to -\infty$) and the mirror at $z=0$ (see Figure 1).  

\begin{figure}
\begin{center}
\includegraphics[width=3.3in]{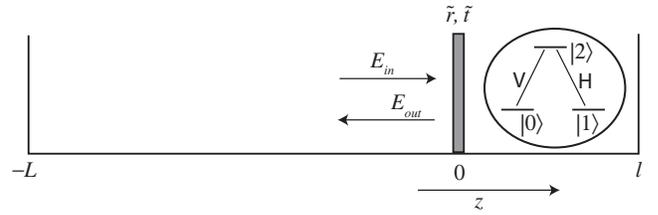}
\end{center}
\caption[example]
   { \label{fig:fig1}
The setup considered in this paper.  The small ($l\ll L$) one-sided cavity has a partly reflecting mirror with amplitude reflection and transmission coefficients $\tilde r$ and $\tilde t$ ($\tilde r^2 + {\tilde t}^2 = 1$), and contains a single three-level system with two degenerate transitions, corresponding (for example) to horizontal and vertical polarizations $H$ and $V$.}
\end{figure}

The mode functions are (Eq.~(2.2) of \cite{universe})
\begin{equation}
U_k(z) = 
\begin{cases}
\xi_k \sin k(z+L) &\text{for $z<0$,} \\
M_k \sin k(z-l) &\text{for $z>0$.}
\end{cases}
\label{e1}
\end{equation}
The $\xi_k$ are taken to be alternately $+1$ and $-1$, and the $M_k$ then are given by
\begin{equation}
M_k = \frac{(c\kappa/l)^{1/2}}{[(\Omega_k-\Omega_c)^2 + \kappa^2]^{1/2}}
\label{e2}
\end{equation}
Here we have changed the notation of \cite{universe} slightly, so that $\kappa$, rather than $\Gamma$, is the cavity amplitude decay rate.  $\Omega_c$ is the central frequency of the cavity quasimode under consideration.  The allowed frequencies $\Omega_k \equiv ck$ and wavevectors $k$ for the modes of the universe can be derived from the eigenvalue equation (Eq.~(A2) of \cite{universe})
\begin{equation}
\tilde r \sin[k(L-l)] = \sin[k(L+l)]
\label{e3}
\end{equation}
where $\tilde r$ is the amplitude reflection coefficient of the cavity input mirror.
The incoming field has the form (Eq.~(2.6a) of \cite{universe})
\begin{equation}
E_\text{in}^{(+)} = \frac{1}{2i} \sum_k \left(\frac{\hbar\Omega_k}{\epsilon_0 A L}\right)^{1/2} \xi_k a_k e^{ikL - i\Omega_k t}
\label{e4}
\end{equation}
and the cavity field is then (Eq.~(2.9) of \cite{universe})
\begin{equation}
E_\text{cav}^{(+)} = \frac{1}{2i} \sum_k \left(\frac{\hbar\Omega_c}{4\epsilon_0 A l}\right)^{1/2} a(t) e^{- i\Omega_c t}
\end{equation}
with the cavity ``single quasimode'' operator
\begin{equation}
a(t) \simeq i \sqrt{\frac l L}\, \sum_k M_k a_k e^{i k l - i (\Omega_k - \Omega_c) t}
\label{e6}
\end{equation}

In this work we actually need to consider two cavity quasimodes, corresponding to horizontal and vertical polarizations.  We will denote the corresponding ``modes of the universe'' operators by $a_{hk}$ and $a_{vk}$, and the quasimode operators by $a_h$ and $a_v$.  We describe the coupling of the atom to the field by the standard interaction-picture Hamiltonian $\hbar g(a \sigma^\dagger + a^\dagger \sigma)$, where $\sigma$ and $\sigma^\dagger$ are the usual atomic lowering and raising operators.  The Hamiltonian then takes the form
\begin{align}
H = &i \hbar g \sum_k M_k^\prime \left(\sigma_{21} a_{hk} e^{- i \delta_k t} - a_{hk}^\dagger \sigma_{12}e^{i \delta_k t} \right) \cr
&+ i \hbar g
\sum_k M_k^\prime \left(\sigma_{20} a_{vk} e^{- i \delta_k t} - a_{vk}^\dagger \sigma_{02}e^{i \delta_k t} \right)
\label{e7}
\end{align}
with 
\begin{equation}
M_k^\prime = \frac{(c\kappa/L)^{1/2}}{[(\Omega_k-\Omega_c)^2 + \kappa^2]^{1/2}} 
\label{e8}
\end{equation}
and $\delta_k = (\Omega_k - \Omega_c) + (\Omega_c - \omega_0)= \Omega_k - \omega_0 = ck - \omega_0$.
Here, both the resonant atomic frequency $\omega_0$ and the coupling constant $g$ are taken to have the same values for the $H$ and $V$ transitions.  This Hamiltonian is to be compared with Eq.~(1) of \cite{koshino}.  The main difference is that we have allowed, through the coefficients $M^\prime_k$, for different coupling strength of the cavity to outside modes with different frequencies, something which is important in the good cavity (small $\kappa$) limit.  

We note that a factor $e^{ikl}$ in (\ref{e6}) has been neglected in writing Eqs.~(\ref{e7}) and (\ref{e8}).  The assumption is that the small cavity is so small that the factor is essentially constant as a function of $k$, that is, if $k=k_c + \Delta k$, then $k_c l \simeq 2 n\pi$ (resonance condition) and $\Delta k l \simeq \kappa l /c \ll 1$ (since, as we shall see later, the range of modes that eventually develop appreciable amplitudes is of the order of $\kappa$ in frequency space).  This assumption may only have to be reevaluated for a very bad cavity; note that in our formalism, $\kappa = c {\tilde t}^2/4 l$, so $\kappa l/c \sim {\tilde t}^2/4$, which under most circumstances should indeed be very small.

Always under the assumption that $l \ll L$, we can take $k = k_c+n\pi/L$, where $n$ is a (positive or negative) integer, and $k_c = \Omega_c/c$.  We do not expect to have more than one photon in the system at any given time, so the state of the system can be written as
\begin{align}
\ket{\psi(t)} = &C_2(t)\ket 2\ket{\text{vac}}_h\ket{\text{vac}}_v + \sum_n C_{hn}(t)\ket 1 {\ket 1}_{hn}\ket{\text{vac}}_v \cr
&+ \sum_n C_{vn}(t)\ket 0 {\ket 1}_{vn}\ket{\text{vac}}_h
\label{e9}
\end{align}
where the first ket refers to the state of the atom, and the second and third to the field; $\ket{\text{vac}}_{h(v)}$ is the vacuum state for the horizontal (vertical) modes; $\ket{1}_{hn}$ is a state with one photon in horizontal mode $n$; and $\ket{1}_{vn}$ is a state with one photon in vertical mode $n$.

Using (\ref{e7}) and (\ref{e9}), one gets the equations of motion
\begin{subequations}
\begin{align}
\dot C_2 &= g \sum_n M^\prime_n C_{hn} e^{-i\delta_n t} + g \sum_n M^\prime_n C_{vn} e^{-i\delta_n t} \\
\dot C_{hn} &= -g M^\prime_n C_2 e^{i\delta_n t} \\
\dot C_{vn} &= -g M^\prime_n C_2 e^{i\delta_n t}
\end{align}
\label{e10}
\end{subequations}
with
\begin{equation}
M_n^\prime = \frac{(c\kappa/L)^{1/2}}{[(n\pi c/L)^2 + \kappa^2]^{1/2}} 
\label{e11}
\end{equation}
and $\delta_n = n\pi c/L + \delta_a$, where $\delta_a = \Omega_c - \omega_0$ is the atom-cavity detuning.

These equations have to be integrated with initial conditions chosen so that the incoming field (\ref{e4}) represents a suitable pulse.  As an example, the set of coefficients below, which we have used for our numerical calculations, describes an incoming Gaussian pulse initially centered at $z_0$ (with $-L < z_0 < 0$), with a duration $T$, that is to say, a spatial width $cT$, and a carrier frequency $\Omega_c + c n_0 \pi/L$.  The possibility of having the initial set of coefficients peak at a frequency different from the cavity frequency is the way to account for a possible detuning between the external field and the cavity in our formalism; here this detuning $\Delta = \Omega - \Omega_c =  c n_0 \pi/L$.   (The analytical calculations in the next Section show how under some conditions one can ``pull out'' $\Delta$, to deal with it in more conventional ways.)
\begin{equation}
C_{n}(0) = \left(\frac \pi 2\right)^{1/4} \sqrt{\frac{cT}{L}} \,e^{-i n\pi z_0/L} \,e^{-(cT (n-n_0)\pi/2 L)^2}
\label{e12}
\end{equation}
Note that the coefficients in (\ref{e12}) are (approximately) normalized to unity.  That is, $\sum_n |C_n(0)|^2 = 1$, provided that the sum extends over a sufficiently large number of modes.  For example, in our numerical work we take $C_{vn}(0) = C_n(0)$, along with $C_2(0)=0$ and all $C_{hn}(0)=0$.  Of course, one could alternately take $C_{hn}(0)=C_n(0)$, with $C_2(0)=0$ and all $C_{vn}(0)=0$.

In principle, a very long initial pulse can be accurately represented by relatively few modes, of the order of a few times $L/cT$; however, it is important to note that in the course of the interaction many more modes, of the order of $\kappa L/c$, may become appreciably excited, as we shall show below.

The formalism introduced here provides a simple framework for numerical calculations, and we shall show the results of some such calculations in Section IV below, but it is worth noting that substantial progress can be made by analytical methods as well.  Accordingly, the following section presents a formal solution of Eqs.~(\ref{e10}) that is quite general (with no further approximations), as well as the simpler results of an adiabatic approximation, valid for sufficiently long pulses.

\section{analytical results}

\subsection{General solution}

Clearly, Eqs.~(\ref{e10}) can be simplified by introducing two sets of variables $C_{n\pm} = (C_{hn}\pm C_{vn})/\sqrt 2$.  Then the $C_{n-}$ are constant, and the system (\ref{e10}) reduces to
\begin{subequations}
\begin{align}
\dot C_2 &= g \sqrt 2 \sum_n M^\prime_n C_{n+} e^{-i\delta_n t} 
\label{e13a}\\
\dot C_{n+} &= -g \sqrt 2 M^\prime_n C_{2} e^{i\delta_n t}
\label{e13b}
\end{align}
\label{e13}
\end{subequations}
That is, the degenerate three level system is formally equivalent to a two-level system, coupled to the polarization $H+V$, plus a ``dark state'' $(\ket 0 - \ket 1)/\sqrt 2$. Hence, we expect our results, in an appropriate limit, to reduce to many results already available for a single two-level system in a cavity, interacting with a single photon pulse \cite{waks1,chen,mei,an}.

We can formally integrate equation (\ref{e13b}) and substitute in (\ref{e13a}).   We get
\begin{align}
\dot C_2 = &-2 g^2 \sum_n {M^\prime_n}^2 \int_0^t e^{-i\delta_n(t-t^\prime)} C_2(t^\prime)\, dt^\prime \cr
&+  g \sqrt 2 \sum_n M^\prime_n C_{n+}(0) e^{-i\delta_n t}
\label{e14}
\end{align}
The second term in (\ref{e14}) is a driving term that depends on the initial condition (that is, the initial shape of the wavepacket); below we shall call it $f_0(t)$ for brevity.  In the first term, we can use
\begin{equation}
{M_n^\prime}^2 = \frac c L \, \frac{\kappa}{(n\pi c/L)^2 + \kappa^2}
\label{e15}
\end{equation}
and
\begin{equation}
\delta_n = \frac{cn\pi}{L} +\delta_a
\label{e16}
\end{equation}
and convert the sum over $n$ into an integral over $\omega$, with $\omega = cn\pi/L$:
\begin{align}
\frac c L &\sum_{n=-\infty}^\infty \frac{\kappa}{(n\pi c/L)^2 + \kappa^2} e^{-i(cn\pi/L)(t-t^\prime)} \cr
&\simeq \frac 1 \pi \int_{-\infty}^\infty \frac{\kappa}{\omega^2 + \kappa^2} e^{-i\omega(t-t^\prime)}\, d\omega \cr
&= e^{-\kappa |t-t^\prime|}
\label{e17}
\end{align}
So now (3) becomes (since $t^\prime \le t$, by construction)
\begin{equation}
\dot C_2 \simeq -2g^2 \int_0^t e^{-(\kappa+i\delta_a)(t-t^\prime)} C_2(t^\prime)\, dt^\prime + f_0(t)
\label{e18}
\end{equation}
with 
\begin{equation}
f(t) = g \sqrt 2 \sum_n M^\prime_n C_{n+}(0) e^{-i\delta_n t}
\label{e19}
\end{equation}
This integro-differential equation is of a form that describes a damped, driven harmonic oscillator and so can be solved exactly.  To make it slightly more general, we can even add a term $-\gamma C_2$ to the right-hand side of (\ref{e18}), to account approximately for spontaneous emission losses.  Such an approach breaks the unitarity of the system (which means the total probability is no longer conserved), and does not properly account for the fact that the atom must return to one of the ground states after a spontaneous emission event, but it should be a reasonable lowest-order approximation when the spontaneous emission probability is small.  

We proceed by taking the derivative of (\ref{e18}), with the optional term $-\gamma C_2$, and using (\ref{e18}) itself to eliminate the integral, with the result
\begin{equation}
\ddot C_2  +(\gamma+\kappa+i\delta_a)\dot C_2 + 2 g^2 +\gamma(\kappa+i\delta_a) = F(t) 
\label{e20}
\end{equation}
with 
\begin{equation}
F(t) = \dot f_0 + (\kappa+i\delta_a)f_0
\label{e21}
\end{equation}
With the initial condition $C_2(0) = \dot C_2(0) = 0$ (atom unexcited before the pulse arrives), the formal solution of this equation is
\begin{equation}
C_2(t) = \frac{1}{\lambda_1-\lambda_2}\int_0^t \left(e^{\lambda_1(t-t^\prime)} - e^{\lambda_2(t-t^\prime)} \right) F(t^\prime)\, dt^\prime
\label{e22}
\end{equation}
in terms of the eigenvalues, $\lambda_1$ and $\lambda_2$, of the characteristic equation
\begin{equation}
\lambda^2 +(\gamma+\kappa+i\delta_a)\lambda + 2 g^2 +\gamma(\kappa+i\delta_a)  = 0
\label{e23}
\end{equation}
We stress that this represents, in principle, a full solution to the problem, valid for any type of cavity (good, bad or in between) and any kind of pulse (fast or slow).  One only needs to use Eq.~(\ref{e19}) (preferably in integral form) to calculate $f(t)$, then obtain $C_2$ from Eqs.~(\ref{e21}--\ref{e23}), then substitute that in Eqs.~(\ref{e13b}), from which one can get all the coefficients of the state (\ref{e9}), and answer any question about the quantum logical operation performed, including its fidelity.  Further simplification of the results is possible, however, in a particularly important case, that of a very slow (adiabatic) pulse, which we consider in the next subsection.

\subsection{Adiabatic approximation for slow pulses}

In this section we consider the case in which the time scale over which the incident pulse changes, given by $T$, is much longer than all the other scales in the problem, and, most importantly, those set by the solutions $\lambda_{1,2}$ of (\ref{e23}):
\begin{equation}
\frac 1 T \ll |\lambda_{1,2}|
\label{e24}
\end{equation}
In this case, things can be simplified, as follows.  In the expression (\ref{e19}), the frequencies $\delta_n$ will be of the form $\delta_n = c(n-n_0)\pi/L + \Delta +\delta_a$, where $c(n-n_0)\pi/L$ is expected to be a slow frequency, and we allow for the possibility of either or both detunings being large (recall $\Delta = n_0 c\pi/L$).  We then write 
\begin{equation}
f(t) = e^{-i(\Delta+\delta_a)t} \tilde f_0(t)
\label{e25}
\end{equation}
where $\tilde f_0$ is a slowly-varying function, and approximate (\ref{e21}) by
\begin{equation}
F(t) \simeq (\kappa - i\Delta)f(t) = (\kappa - i\Delta) e^{-i(\Delta+\delta_a)t} \tilde f_0(t)
\label{e26}
\end{equation}
We can then use an adiabatic approximation to the integral (\ref{e22}), which consists in integrating by parts and neglecting the second term (which in turn can be used to estimate the error in the approximation, as will be shown later):
\begin{align}
\int_0^t & e^{-(\lambda+i(\Delta+\delta_a))t^\prime} \tilde f_0(t^\prime)\, dt^\prime =-\frac{e^{-(\lambda+i(\Delta+\delta_a))t}}{\lambda+i(\Delta+\delta_a)}\,\tilde f_0 \Bigg|_0^t \cr
&+\frac{1}{\lambda+i(\Delta+\delta_a)}\int_0^t e^{-(\lambda+i(\Delta+\delta_a))t^\prime} \dot{\tilde f}_0(t^\prime)\, dt^\prime
\label{e27}
\end{align}
The result is then
\begin{align}
C_2 &\simeq \frac{\kappa - i\Delta}{(\lambda_1+i(\Delta+\delta_a))(\lambda_2+i(\Delta+\delta_a))}\, f_0(t) \cr
&= \frac{\kappa - i\Delta}{2g^2-(\kappa-i\Delta)(\gamma+i(\Delta+\delta_a))}\, f_0(t)
\label{e28}
\end{align}
Finally, this can be substituted back in Eq.~(\ref{e13b}), which, using the explicit definition (\ref{e19}) for $f_0$, yields
\begin{align}
\dot C_{n+} = -&\frac{2 g^2(\kappa - i\Delta)}{2g^2-(\kappa-i\Delta)(\gamma+i(\Delta+\delta_a))} \cr 
&\times\sum_m  M^\prime_n M^\prime_m C_{m+}(0) e^{-i(\delta_m -\delta_m) t}
\label{e29}
\end{align}
The simplest way to handle (\ref{e29}) is to \emph{formally} integrate the right-hand side all the way to the large cavity roundtrip time (the ``quantization time'') $2L/c$, since in that case all the terms in the sum vanish except the one with $m=n$.  This may seem unphysical, since, in the limit $L\to \infty$, the roundtrip time becomes infinite; but in fact what we have on the right-hand-side of (\ref{e29}) is just something proportional to $e^{i\delta_n t} f_0(t)$, and the integral of this from 0 to $t_\text{max}$ will assume its final value as soon as $t_\text{max}$ comfortably exceeds the duration of the pulse $T$, at which point there is no harm in letting $t_\text{tmax}$ formally go to infinity or, in the discrete-mode picture, to the roundtrip time $2L/c$.  Under those conditions, we get, for sufficiently large $t$,
\begin{widetext}
\begin{align}
C_{n+}(t_\infty) &= C_{n+}(0) -\frac{2 g^2(\kappa - i\Delta)}{2g^2-(\kappa-i\Delta)(\gamma+i(\Delta+\delta_a))}  {M_n^\prime}^2 \frac{2L}{c}\, C_{n+}(0) \cr
&= C_{n+}(0) \left [1 - \frac{2 g^2(\kappa - i\Delta)}{2g^2-(\kappa-i\Delta)(\gamma+i(\Delta+\delta_a))}  \,\frac{2\kappa}{(nc\pi/L)^2 + \kappa^2}\right]
\label{e30}
\end{align}
Again, for a very long pulse ($\kappa T \gg 1$) it is consistent to replace $nc\pi/L$ in (\ref{e30}) by $\Delta = n_0c\pi/L$, in which case, after some algebra, the final result can be written as
\begin{equation}
C_{n+}(t_\infty) = -\frac{2g^2(\kappa-i\Delta) + (\kappa^2+\Delta^2)(\gamma+i(\Delta+\delta_a))}{2g^2(\kappa+i\Delta) - (\kappa^2+\Delta^2)(\gamma+i(\Delta+\delta_a))} \, C_{n+}(0)
\label{e31}
\end{equation}
Note that when $g=0$ (the pulse does not couple to the atom; for instance, if it has the $H$ polarization when the initial state of the atom is zero) one just gets $C_{n+}(t_\infty) = C_{n+}(0)$.  

The result (\ref{e31}) can also be written with an explicit phase factor pulled out:
\begin{equation}
C_{n+}(t_\infty) = -\frac{\kappa-i\Delta}{\kappa+i\Delta}\times\frac{2g^2 + (\kappa+i\Delta)(\gamma+i(\Delta+\delta_a))}{2g^2 - (\kappa-i\Delta)(\gamma+i(\Delta+\delta_a))} \, C_{n+}(0)
\label{e32}
\end{equation}
In this form, the second factor is directly comparable to the reflection coefficient calculated by Waks and Vuckovic \cite{waks1}, and used in various limits by many other authors \cite{chen,mei,an}.  The reason that the prefactor appears here but not in the reflection coefficient is discussed in the next section.
\end{widetext}

\subsection{Comparison with previous results}
Equation (\ref{e31}) clearly shows that when spontaneous emission is negligible, the final state of the system consisting of a single-photon pulse with the ``$+$'' polarization and the atom in the ``$+$'' ground state is merely a phase factor times the initial state.  This effect has been considered before (it is sometimes referred as a ``Faraday effect'') and quantum logic based on it has also been proposed \cite{chen,mei,an}.  Any other initial states of the field or of the atom, can always written as superpositions of ``$+$'' and ``$-$'' states, from which it follows that the most general result one can have is a rotation of the polarization of the photon, accompanied by a rotation of the state of the atom in the $\ket 0,\ket 1$ basis.  The interesting thing is that in general this is an entangling operation.

More specifically, let $\gamma = 0$.  We can write 
\begin{equation}
C_{n+}(t_\infty) = - e^{2i\phi}C_{n+}(0)
\label{e33}
\end{equation}
with
\begin{equation}
\phi = \tan^{-1}\left[\frac{(\Delta+\delta_a)(\kappa^2+\Delta^2)-2g^2\Delta}{2 g^2\kappa}\right]
\label{e34}
\end{equation}
We also know that $C_{n-} \equiv (C_{hn}-C_{vn})/\sqrt 2$ does not change with time.  We therefore have the transformation
\begin{align}
C_{hn}+C_{vn} &\to -e^{2i\phi} \left(C_{hn}(0) + C_{vn}(0)\right) \cr
C_{hn}-C_{vn} &\to  C_{hn}(0) - C_{vn}(0)
\label{e35}
\end{align}
Adding and subtracting these expressions, we find, in the notation of Koshino et al., the following evolution for the four possible initial basis states:
\begin{align}
\ket{H,0} &\to \ket{H,0} \cr
\ket{H,1} &\to -e^{i\phi}\left[i \sin\phi \ket{H,1} + \cos\phi\ket{V,0} \right] \cr
\ket{V,0} &\to -e^{i\phi}\left[\cos\phi \ket{H,1} + i\sin\phi \ket{V,0} \right] \cr
\ket{V,1} &\to \ket{V,1} 
\label{e36}
\end{align}
When $\phi=0$ one has essentially (up to a sign) a SWAP gate between the photon and the atom, whereas when $\phi=\pi/4$ one has a $\sqrt{\text{SWAP}}$, as pointed out in \cite{koshino}.  To have $\phi = 0$ it is sufficient that all the detunings vanish.  The condition to have $\phi=\pi/4$ is, by (\ref{e34}),
\begin{equation}
(\Delta+\delta_a)(\kappa^2+\Delta^2)-2 g^2(\Delta+\kappa) = 0
\label{e37}
\end{equation}
which can always be satisfied, in principle.  An important insight of Koshino et al. is that not only is the $\sqrt{\text{SWAP}}$ a universal gate for quantum computation, but by applying it twice one can get an ordinary SWAP as well.  Hence a $\sqrt{\text{SWAP}}$ gate between two photonic qubits can be carried out in a completely passive way, without any need to directly manipulate the atom, the cavity, or the photon detuning, simply by reflecting the first pulse twice successively from the cavity, which swaps its state for that of the atom; reflecting the second pulse, which carries out a $\sqrt{\text{SWAP}}$ between the second photon and the atomic qubit; and finally reflecting the first photon twice again, so that it acquires the state of the atom after the $\sqrt{\text{SWAP}}$.

The specific results of Koshino et al. can be derived easily from (\ref{e31}) by setting $\gamma = \delta_a=0$ and taking the bad cavity limit: $\kappa \gg g, \Delta$.  Introducing $\Gamma = 2 g^2/\kappa$, one has then $C_{n+}(t_\infty) \simeq -C_{n+}(0)(\Gamma + i\Delta)/(\Gamma+i\Delta)$, and the condition for $\phi=\pi/4$ is just $\Delta = \Gamma$.  One of the key results of our analysis is that the $\sqrt{\text{SWAP}}$ gate can be carried out in a much broader range of regimes, the only necessary conditions being that the pulse be very long (adiabatic condition) and that spontaneous emission be negligible.

As pointed out in the previous section, many previous studies based on the standard input-otput formalism \cite{collet} for the field operators make use of a reflection coefficient given by the second factor in Eq.~(\ref{e32}), and infer from it the phase change of the total atom-field state upon reflection.  In our formalism, as well as in \cite{koshino}, the total phase factor for the state, as written in (\ref{e32}), equals this ``reflection coefficient'' multiplied by a prefactor.  We can show that if one's specific goal is to compare expectation values of the input field operator to expectation values of the output field operator, the prefactor cancels, as follows: by Eq.~(2.6b) of \cite{universe}, the output field operator is
\begin{equation}
E_\text{out}^{(+)} = -\frac{1}{2i} \sum_k \left(\frac{\hbar\Omega_k}{\epsilon_0 A L}\right)^{1/2} \xi_k a_k e^{-ikL - i\Omega_k t}
\label{e38}
\end{equation}
The difference between this and the input field operator, Eq.~(\ref{e4}), is a factor $-e^{2ikL}$ in every term in the sum.  Although one may expect this to be very close to $-1$ for all relevant values of $k$, this approximation is generally too drastic, and one needs to use the eigenvalue equation (\ref{e3}) for a better estimate.  By writing all the trigonometric functions as sums of complex exponentials, it is easy to see that (\ref{e3}) is equivalent to
\begin{equation}
e^{-2ikL} = \frac{\tilde r e^{-ikl} - e^{ikl}}{\tilde r e^{ikl} - e^{-ikl}} \simeq \frac{-2 i \sin kl - (\tilde t^2/2) e^{-ikl}}{2 i \sin kl - (\tilde t^2/2) e^{ikl}}
\label{e39}
\end{equation}
where the approximation $\tilde r \simeq 1- \tilde t^2/2$ has been used in the last step.  If we then assume $kl = 2n\pi +\Delta l/c$, where $\Delta$, as above, is the incoming field detuning from the cavity resonance, and expand on small quantities, we can conclude that
\begin{equation}
E_\text{out}^{(+)} = -\frac{\kappa + i\Delta}{\kappa - i\Delta}\, E_\text{in}^{(+)}
\label{e40}
\end{equation}
In obtaining this result, we used the relationship $\kappa = c{\tilde t}^2/4 l$ between $\kappa$ and the mirror transmission coefficient $\tilde t$.  Equation (\ref{e40}) shows that if the state coefficients (\ref{e32}) are used to calculate the expectation values of the output field, the prefactor (including the overall minus sign) will cancel, and the result will be related to the expectation values of the input field by only the second factor in (\ref{e32}), that is, the ``reflection coefficient'' of the input-output formalism.  Conversely, note that for reflection off of an empty cavity (or one where the atom does not couple to the field, $g=0$), the input-output formalism predicts a phase shift of precisely $-(\kappa + i\Delta)/(\kappa - i\Delta)$ (whereas the present formalism yields no change in the state vector coefficients); hence, in the input-output formalism, the right-hand sides of Eq.~(\ref{e36}) would all be consistently multiplied by a phase factor $-(\kappa + i\Delta)/(\kappa - i\Delta)$, which, of course, makes no difference physically.  Despite this agreement, note that, in general, the approach of ascribing to the Schr\" odinger-picture states the phase shifts derived for the Heisenberg-picture field operators works only when there is only one photon in the field.  When dealing with coherent states, for example, there is a big difference between the state $\ket{-\alpha}$ and the state $-\ket{\alpha}$.

\section{numerical results}

In this section we show some results of the numerical integration of the system (\ref{e13}) in order to examine the validity of the adiabatic approximation and to characterize the effect of spontaneous emission on the operation of the $\sqrt{\text{SWAP}}$ gate.  Alternatively, one could, in principle, use instead the exact (in the continuous limit) solution (\ref{e22}), which, for the Gaussian pulse we will consider, can be evaluated in terms of error functions of complex arguments; but, after that, the algebra becomes cumbersome, whereas the direct numerical integration of Eqs.~(\ref{e13}) poses no particular challenges.  It is, however, necessary to realize that, even if the initial pulse has a very small bandwidth, once it is inside the cavity this changes, and one may need to consider a very large number of ``modes of the universe,'' of the order of $\kappa L/c$ or $\Gamma L/c$, in order to accurately determine the asymptotic behavior investigated below.  This point is elaborated in more detail towards the end of this Section.

We take the spectrum of the initial pulse to be given by the coefficients (\ref{e12}), out of which we can build the initial coefficients $C_{n+}(0)$, which will formally be equal to $C_n/\sqrt 2$. Then $C_{n-}(0)=\pm C_n/\sqrt 2$, depending on whether we take this to be a vertically or horizontally polarized pulse.  In any case, the $C_{n-}(0)$ are constant, so for a measure of the performance of the gate it suffices to consider the $C_{n+}(0)$.  In particular, we shall define
\begin{equation}
F e^{2i\Phi} = -\frac{\sum_n C_{n+}^\ast(0) C_{n+}(\infty)}{\sum_n C_{n+}^\ast(0) C_{n+}(0)} = -2\sum_n C_{n+}^\ast(0) C_{n+}(\infty)
\label{e41}
\end{equation}
where $F$ is a measure of the ``fidelity'' of the final wavepacket to the initial one, and $\Phi$ measures the phase shift of the state in the Schr\" odinger picture.  Comparing to Eq.~(\ref{e33}) we see that if the adiabatic approximation were exact, one would have $F=1$ and $\Phi = \phi$ as given by Eq.~(\ref{e34}); the deviation from these values can be used to characterize the gate error.

To begin with, we show in Fig.~2 that the gate works well in both the bad cavity and good cavity limits, provided only that the pulse is sufficiently long. This requires both $T\gg 1/\kappa$ and $T \gg 1/\Gamma$ (with $\Gamma\equiv 2g^2/\kappa$).  For the parameters of the figure, the first inequality is always satisfied, whereas the second one requires $g/\kappa \gg 0.07$.  Note that this includes both bad cavity ($g < \kappa$) and good cavity ($g > \kappa$) values.

\begin{figure}
\begin{center}
\includegraphics[width=3.3in]{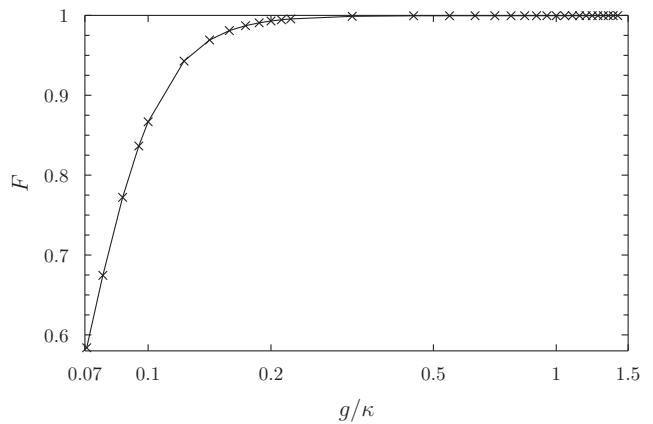}
\end{center}
\caption[example]
   { \label{fig:fig2}
Fidelity $F$ versus coupling constant $g$ for an initial Gaussian pulse of duration $T=100/\kappa$.  For every value of $g/\kappa$, $\Delta/\kappa$ is chosen so as to satisfy the condition Eq.~(\ref{e37}), with $\delta_a=0$.}
\end{figure}

The way the gate fidelity improves as the pulse duration increases is illustrated in Figure 3, for $\Gamma = 0.5\kappa$ (i.e., $g=\kappa/2$) and variable $T$.  The best fit slope in the log-log plot is $-1.99$, which suggests that the fidelity improves quadratically as $T$ increases.

\begin{figure}
\begin{center}
\includegraphics[width=3.3in]{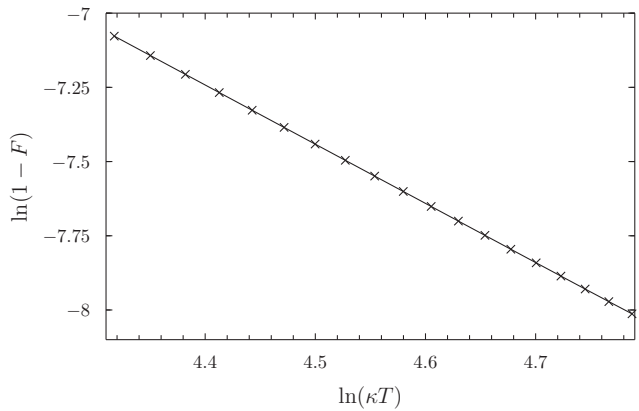}
\end{center}
\caption[example]
   { \label{fig:fig3}
``Infidelity'' $1-F$ versus pulse duration $T$ for $g=\kappa/2$, and $\Delta$ and $\delta_a$ as in Fig.~2.}
\end{figure}

This quadratic dependence may be a little surprising, since a glance at Eq.~(\ref{e27}) suggests that the terms neglected in the adiabatic approximation only decay as $1/T$.  This is true in general, but a closer examination reveals that for a long pulse with a symmetric frequency spectrum, the more favorable $1/T^2$ can be approximately realized.  To see this, note that an adiabatic approximation to the second term in (\ref{e27}) yields essentially $e^{-i(\Delta+\delta_a)t}\dot{\tilde f}_0(t)/[\lambda+i(\Delta+\delta_a)]^2$, where by Eq.~(\ref{e25}) $\dot{\tilde f}_0$ goes as
\begin{equation}
\dot{\tilde f}_0(t) = i e^{i(\Delta+\delta_a) t} \sum_n M_n^\prime C_{n+}(0)\left(\Delta+\delta_a - \delta_n\right) e^{i\delta_n t}
\label{e42}
\end{equation}
When this is substituted in Eq.~(\ref{e13b}) and the integration over the quantization time is performed, one obtains a correction to $C_{n+}$ which goes as
\begin{align}
\delta C_{n+}(\infty) &\propto \frac{\Delta+\delta_a-\delta_n}{(nc\pi/L)^2 + \kappa^2} C_{n+}(0) \cr
&\propto  \frac{n-n_0}{(nc\pi/L)^2 + \kappa^2} C_{n+}(0)
\label{e43}
\end{align}
(making use of the fact that $\delta_n = c(n-n_0)\pi/L + \Delta +\delta_a$, as indicated below Eq.~(\ref{e24})).  For a pulse such as (\ref{e12}), for which $|C_n|^2$ is an even function of $n-n_0$, the sum $\sum_n \delta C_{n+}(\infty) C_{n+}^\ast(0)$ will vanish, which means that the next-order contribution to the infidelity will come from $\ddot{\tilde f}_0 \sim 1/T^2$, in agreement with the numerical results.  Figure 4 shows that this also applies to the gate phase error, that is to say, the difference between $\Phi$ and the value $\pi/4$ needed for the $\sqrt{\text{SWAP}}$ gate also appears to decrease at least quadratically, although here the results of the numerical integration are not so clean.

\begin{figure}
\begin{center}
\includegraphics[width=3.3in]{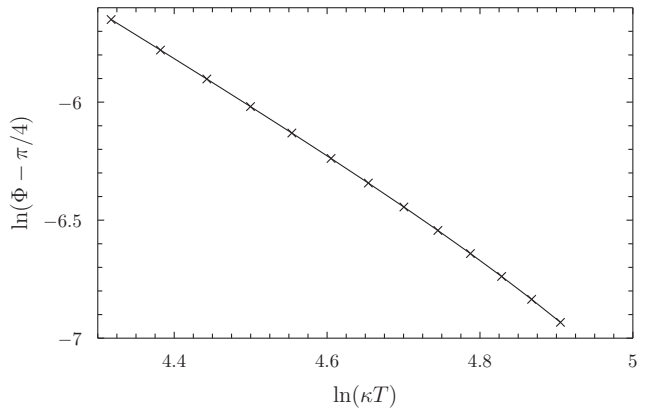}
\end{center}
\caption[example]
   { \label{fig:fig4}
Difference between the phase factor $\Phi$ and $\pi/4$ versus pulse duration $T$ for the same parameters as in Fig.~3. The best straight-line fit has slope  $-2.17$.}
\end{figure}

The theoretical possibility to carry out a deterministic gate between single photons with an error scaling as the inverse square of the gate duration is remarkable in view of the results derived by one of us in \cite{Gea03} for quantum logic with atomic qubits.  There it was shown that if the gates were mediated by a quantized field in a coherent state, with an average number of photons $\bar n$, the gate error would scale as $1/\bar n$.  Note that this constraint would apply to the original Duan-Kimble gate \cite{DandK}, which requires the manipulation of the atom by a coherent field in between single-photon pulse reflections; in contrast, the Koshino-Ishizaka-Nakamura gate appears to be able to work with negligible error at the single-photon level.  

The $1/T^2$ scaling derived here is also better than the $1/T$ scaling postulated to hold, alternatively, in \cite{Gea03} for quantum logic with material qubits and quasistatic control fields.  Time and again over the past few years one or the other of these constraints has been shown to hold for all kinds of systems of experimental interest \cite{mayo}, and in many cases it has been shown to follow from very fundamental considerations, such as spontaneous emission \cite{geareply}, or conservation laws \cite{ozawa1,geaozawa1}.  The present result, however, appears to suggest that quantum logic should be possible with a much smaller energy cost than had previously been thought possible (although one cannot yet rule out the possibility of some hidden energy cost in, for instance, the electronic devices that might be used to route the photons towards or away from the cavities).  In any case, this is a question that certainly requires further investigation. 

Besides the breakdown of the adiabatic approximation, an important factor that may degrade the performance of the gate is spontaneous emission.  Figures 5 and 6 compare the fidelities and phase shifts obtained numerically for zero and nonzero $\gamma$, and show that the damage done by $\gamma$ is approximately linear for the fidelity and quadratic for the phase angle.  These scalings agree with the result of a simple expansion in powers of $\gamma$ of the adiabatic approximation Eq.~(\ref{e31}).

\begin{figure}
\begin{center}
\includegraphics[width=3.3in]{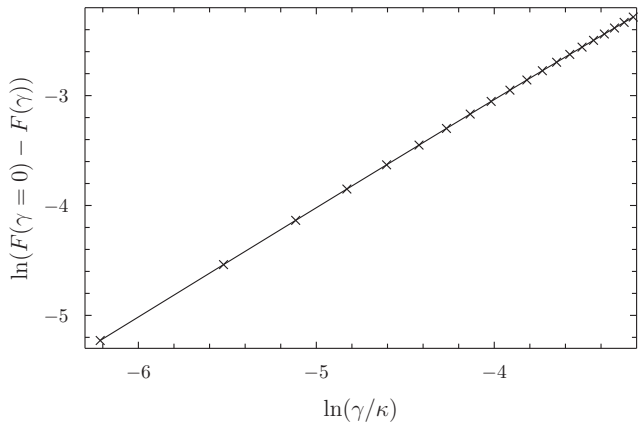}
\end{center}
\caption[example]
   { \label{fig:fig5}
Difference between the fidelity obtained numerically when $\gamma=0$ and when $\gamma$ is nonzero.  The pulse duration $T=100/\kappa$, and other parameters are as in Figs.~3 and 4.  The best-fit slope is $0.98$.}
\end{figure}

\begin{figure}
\begin{center}
\includegraphics[width=3.3in]{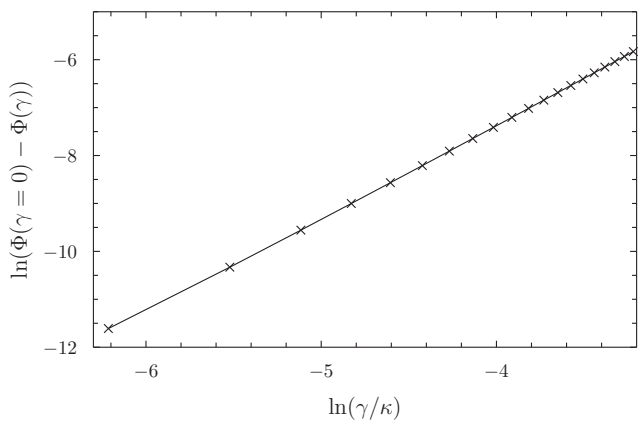}
\end{center}
\caption[example]
   { \label{fig:fig6}
Difference between the phase factor $\Phi$ obtained numerically when $\gamma=0$ and when $\gamma$ is nonzero.  Parameters as in Fig.~5.  The best-fit slope is $1.94$.}
\end{figure}

We note that our treatment of spontaneous emission is formally equivalent to considering only the first term in a ``stochastic wavefunction'' unraveling of the underlying master equation \cite{carmichael,molmer,hegerfeldt,dum}. Up to an overall normalization factor, this just gives the evolution of the system conditioned on the probability of no actual spontaneous emission event taking place.  The probability to, in fact, lose the photon because of spontaneous emission can be estimated from the adiabatic approximation as
\begin{align}
P_\text{loss} &= 2\gamma \int_0^\infty |C_2(t)|^2\, dt \cr
&= \frac{4\gamma g^2\kappa}{(2 g^2 - \Delta(\Delta + \delta_a))^2 + \kappa^2(\Delta+\delta_a)^2}
\label{e44}
\end{align}
Making use of the condition (\ref{e37}) which the detunings have to satisfy in order to have a $\sqrt{\text{SWAP}}$ gate, this simplifies substantially, to
\begin{equation}
P_\text{loss} = \frac{\kappa \gamma}{2 g^2}\left[1+\left(\frac \Delta \kappa\right)^2\right]
\label{e45}
\end{equation}
which suggests that the optimal detuning arrangement, to minimize $P_\text{loss}$, is to have $\Delta = 0$ (pulse resonant with the cavity) and $\delta_a = 2 g^2/\kappa$ (atom detuned from both the cavity and the pulse).  In that case, the probability that the photon be lost through spontaneous emission is just the inverse of the ``Purcell factor,'' $F=2g^2/\kappa\gamma$.

Finally, it is also possible to use the numerical results for $C_{n+}(t)$ to visualize the evolution of the pulse in space, as well as in the frequency domain.  Figure 7 shows the pulse intensity $I_\text{in}(t,z)=\av{E_\text{in}^{(-)}(t,z)E_\text{in}^{(+)}(t,z)}$ at three instants, before, during and after the interaction with the cavity; the initial field-atom state is taken to be $\ket{V,0}$, and the detuning is taken to satisfy the condition for the $\sqrt{\text{SWAP}}$, in which case Eq.~(\ref{e36}) predicts, and the numerical calculations show, that the initial pulse intensity is reduced by a factor of 2 and a horizontally polarized pulse with the same reduced intensity is generated by the interaction.  Note that, by Eq.~(\ref{e38}) and the discussion following it regarding the phase factors $e^{2ikL}$, the outgoing field intensity $I_\text{out}(t,z)$ is formally identical to $I_\text{in}(t,-z)$, so the part of Figure 7 corresponding to $z>0$ can be regarded as an ``unfolded'' (reflected around the $z=0$ plane) view of the evolution of the outgoing pulse in the one-sided cavity setup of Figure 1.  This also means that the cavity itself is invisible in the figure; the vertical line at $z=0$ merely represents its input (which is also its output) mirror.

\begin{figure}
\begin{center}
\includegraphics[width=3.3in]{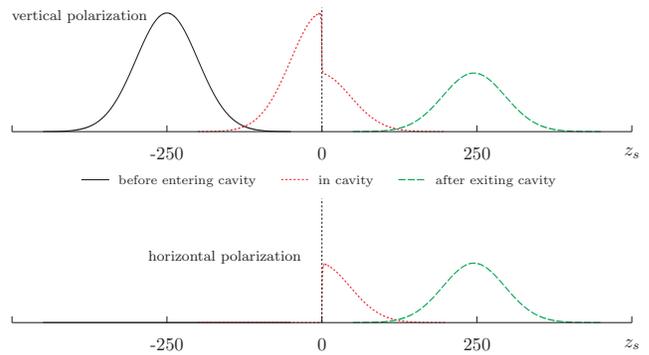}
\end{center}
\caption[example]
   { \label{fig:fig7}
Pulse intensity profiles, at three different times, for the case of a vertically-polarized pulse incident on the cavity under $\sqrt{\text{SWAP}}$ conditions. The reflected pulse has both horizontal- and vertically-polarized components.  The position $z$ has been scaled by the distance that the pulse travels in a cavity lifetime, that is, $z_s = z\kappa/c$.  The dotted vertical line represents the position of the cavity and the view is ``unfolded'' so that $z_s>0$ corresponds to a reflected pulse (see text for details).  The parameters for the calculation are $\Gamma = 0.1\kappa$, $T=100/\kappa$.}
\end{figure}

The need for the spectral coefficients $C_{n+}$ to reproduce the sharp discontinuity exhibited in Fig. 7 by the analytical functions $I_h(t,z)$ and $I_v(t,z)$ explains why it is necessary to keep a large number of coefficients in the calculation for accuracy.  Figure 8 shows that the spectra themselves are clearly broadened somewhat when the pulse is in the cavity; in particular, a close scrutiny shows that there are long, but relatively low magnitude, tails that cover the whole cavity bandwidth.  Figure 8 also shows that the $\sqrt{\text{SWAP}}$ condition, here satisfied by $\delta_a=0$ and $\Delta \simeq \Gamma$, results in some frequency pulling of the spectrum towards the cavity line center (at $n=0$, or $n-n_0 = -n_0$), while the pulse is in the cavity.

\begin{figure}
\begin{center}
\includegraphics[width=3.3in]{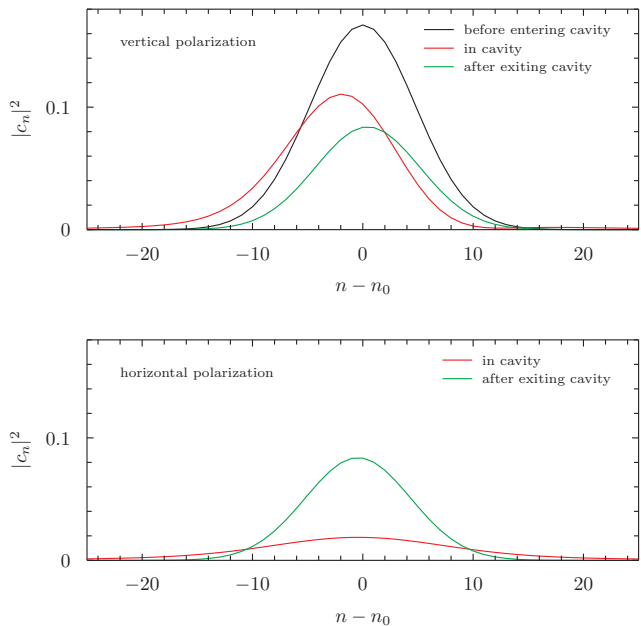}
\end{center}
\caption[example]
   { \label{fig:fig8}
Frequency spectra of the different polarizations for the situation depicted in Fig. 7.}
\end{figure}

\section{Conclusions}

In this paper, we have extended the analysis of \cite{koshino} to show that the kind of all-passive, cavity-mediated, single photon quantum logic proposed by Koshino, Ishizaka and Nakamura actually can work in the good cavity limit as well as in the bad cavity limit originally considered in \cite{koshino}.  We have derived formally exact analytical expressions for the excitation probability, and approximate results valid for sufficiently slow pulses.  We have shown that the gate error probability in the absence of spontaneous emission scales as $1/T^2$, where $T$ is the pulse duration, and given expressions for the photon-loss probability due to spontaneous emission valid also in the adiabatic limit.  Our formalism also has allowed us to clarify the relationship between the phase factor for the reflected pulse derived by Koshino et al. and previous results for two-level systems found in the literature.

We believe that the possibility of deterministic quantum logical gates at the single-photon level, without the need for complicated manipulations, opened up by the work in \cite{koshino}, deserves a great deal of attention.  Many groups are currently working on possible schemes for all-optical quantum computers \cite{kok} that suffer from enormous overheads because of the probabilistic nature of available photon-photon logical gates.  Even if an all-optical quantum computer turned out not to be the best choice in the long run, deterministic gates for single photons could still prove invaluable for quantum communication, to perform, for instance, entanglement distillation or state purification on photons carrying quantum information across nodes of a quantum network.  Further studies are clearly necessary to identify suitable systems, and to characterize the performance of these types of gates in realistic setups with currently-available technology.  

Finally, as we have indicated in the previous section, the possibility of these gates also raises questions of fundamental interest, such as the true minimum energy requirements for quantum computation, which we intend to pursue in the near future.

\end{document}